\newcommand\kms{$\rm\,km\,s^{-1}\,$}
\newcommand\kmsm{$\rm\,km\,s^{-1}Mpc^{-1}\,$}
\newcommand\etal{{\it et al.}\ }

\documentstyle[11pt,aaspp4]{article}
\slugcomment{(to appear in the Astronomical Journal)}

\lefthead{Scodeggio, Giovanelli \& Haynes}
\righthead{Universality of the Fundamental Plane. Photometry}

\begin{document}

\title{The Universality of the Fundamental Plane of E and S0 Galaxies.\\
Sample Definition and I Band Photometric Data}

\author{Marco Scodeggio\altaffilmark{1,2}, Riccardo
Giovanelli\altaffilmark{1} and Martha P. Haynes}
\affil{Center for Radiophysics and Space Research and National Astronomy
and Ionosphere Center\altaffilmark{3}, Cornell University, Ithaca, NY 14853\\
mscodegg@eso.org \\ riccardo, haynes@astrosun.tn.cornell.edu}

\altaffiltext{1}{Visiting Astronomer, Kitt Peak National Observatory, 
National Optical Astronomy Observatories, which is operated by the Association 
of Universities for Research in Astronomy, Inc. (AURA) under cooperative 
agreement with the National Science Foundation.}
\altaffiltext{2}{Present address: European Southern Observatory, 
Karl-Schwarzschild-Stra$\beta$e 2, D-85748, Garching bei M{\"u}nchen, 
Germany.}
\altaffiltext{3}{The National Astronomy and Ionosphere Center is
operated by Cornell University under a cooperative agreement with the
National Science Foundation.}

\begin{abstract}
As part of a project to compare the Fundamental Plane and Tully-Fisher
distance-scales, we present here I band CCD photometry for 636
early-type galaxies in 8 clusters and groups of galaxies. These are
the A262, A1367, Coma (A1656), A2634, Cancer and Pegasus clusters, and
the NGC 383 and NGC 507 groups.  Sample selection, cluster properties,
and cluster membership assignment criteria are discussed. We present
photometric parameters that are used in the Fundamental Plane
relation, the effective radius $r_e$ and the effective surface
brightness $\mu_e$, as derived from a $r^{1/4}$ fit to the observed
radial photometric profile of each galaxy.  A comparison with similar
data found in the literature for the Coma cluster shows that large
systematic uncertainties can be introduced in the measurement of $r_e$
and $\mu_e$ by the particular method used to derive those parameters.
However the particular combination of these two parameters that enters
in the Fundamental Plane relation is a quantity that can be measured
with high accuracy.
\end{abstract}

\keywords{distance scale --- galaxies: distances and
redshifts --- galaxies: elliptical and lenticular --- galaxies:
fundamental parameters --- galaxies: photometry}

\section{Introduction}

This is the third in a series of papers (Scodeggio, Giovanelli \&
Haynes 1997a,b; hereafter Paper I and Paper
II\markcite{pap1}\markcite{pap2}) devoted to a study of the
universality of the Fundamental Plane (FP) relation of early-type
galaxies (Djorgovski \& Davis 1987\markcite{DD}; Dressler \etal
1987\markcite{Dre87}). This relation, and the equivalent Tully-Fisher
(TF) relation for spiral galaxies (Tully \& Fisher 1977\markcite{TF}),
have been widely used in the recent past to produce large samples of
redshift-independent distance estimates.  The small scatter of the FP
relation has been also used to put constraints on the epoch of
early-type galaxy formation, and on the subsequent evolution and star
formation histories of those objects (see for example Guzm{\`a}n,
Lucey \& Bower 1993\markcite{Guz93}; Renzini \& Ciotti
1993\markcite{Renz93}; van Dokkum \& Franx 1996\markcite{vDo96}; Pahre
\& Djorgovski 1997\markcite{Pahre97}).

Both applications are based on the assumption that the FP and TF
relations are universal: all galaxies in all locations and
environments have effective radius, effective surface brightness and
velocity dispersion, or magnitude and rotation velocity that are
related by the same relation, within measurement uncertainties and the
intrinsic dispersion present in the relations.  However, this might not
be the case, since a number of environmental effects are known to
exist that have the potential of altering significantly the stellar
population and star formation history of the affected
galaxies. Therefore, the FP and TF universality should not be taken for
granted, but should rather be observationally verified (discussions on
this subject are presented by Djorgovski, de Carvalho
\& Han 1988\markcite{Djorg88}; Guzm{\'a}n \etal 1992\markcite{Guz92}; 
J{\o}rgensen, Franx \& Kj{\ae}rgaard 1996\markcite{Jor96}; Schroeder
1996\markcite{Schro96}; Giovanelli \etal 1997b\markcite{G97b}; and in Paper I).

The most direct way to test the reliability of a distance indicator
relation is to compare its predictions with those of another
independent, and at least equally reliable, method. The TF and FP
relations provide the opportunity for such a comparison. They have
similar accuracy, and they are independent methods, because they are
applied to galaxies of different morphological type and different
clustering properties.  Because of the limited accuracy of the
distance estimates obtained with the TF and FP relation, a direct
comparison of the two methods can be done most effectively using
clusters of galaxies, where large samples can be used to reduce the
statistical uncertainty in the estimate of the cluster distances.  It
should be kept in mind, however, that given the existing correlation
between cluster richness and the properties of the intracluster medium
(see for example Edge \& Stewart 1991\markcite{Edge91}), it is
desirable that such a comparison bridge a set of clusters
characterized by a wide range of richness, in order for the TF-FP
comparison to have broad validity.

In a recent work, Giovanelli \etal (1997a,b; hereafter
G97a,b\markcite{G97a}\markcite{G97b}) have presented TF measurements
for a large sample of cluster objects.  This represent the largest
dataset currently available that can be used for estimating cluster
distances using the TF relation.  In this and the following papers of
this series, we present the results of a FP study of a subset (eight
clusters) of the clusters studied by G97a,b. The main purpose of this
work is to obtain distance estimates for those clusters, with
accuracies comparable to those achievable using the TF dataset, to be
used for a comparison between the two methods. In this paper, we
briefly describe the cluster sample selection and the adopted cluster
membership criteria, and we present I band CCD photometry for 636
early-type galaxies in those clusters.  In a companion paper
(Scodeggio, Giovanelli \& Haynes 1998a\markcite{pap4}), we present the
results of spectroscopic observations of a subset of these
galaxies. Following that work, we will discuss the properties of the I
band FP relation, and the comparison between TF and FP distance
estimates for the clusters in our sample (Scodeggio, Giovanelli \&
Haynes 1998b\markcite{pap5}).

The cluster sample selection is described in Section 2; observations
and the data reduction procedure are described in Section 3 and 4; the
photometric sample is presented in Section 5. An assesment of the
quality of the photometric data presented in this work is presented in
Section 6.  Throughout the paper we parametrize distance--dependent
quantities via H$_\circ=100h$ \kmsm.

\section{Cluster sample selection and membership}

Recently G97a,b have presented TF measurements for a sample of 
$\simeq 800$ galaxies in 24 clusters. Because one of the goals of that
work was to obtain a kinematical rest frame that could be used to
study the local peculiar velocity field, it was important that such 
sample of clusters had a rather uniform sky coverage, so that localized
anomalies in the velocity field would average out in the global
sample.  On the contrary, the main goal of this work is a detailed,
point-by-point comparison of distance estimates, irrespective of the
possible peculiar velocities that the target clusters might
have. Therefore a uniform sky coverage is not an important requirement
for our sample. It is essential instead to cover a wide spread in
cluster richness and the properties of the intracluster medium (ICM),
since known environmental effects on galaxies are strongly dependent
on the local galaxy density, or on the density of the ICM. We have
therefore selected a subset of the clusters in G97a,b as targets for
the FP study, restricted to the northern hemisphere, mainly to
facilitate the spectroscopic observations.  The set is composed of the
8 clusters and rich groups of galaxies NGC 383 group, NGC 507 group,
A262, Cancer, A1367, Coma (A1656), Pegasus, A2634.

\subsection{Cluster properties}

Table 1 lists the most important properties of these clusters and
groups.  The cluster center is generally assumed to be coincident with
the position of the brightest galaxy, except in the case of the
irregular cluster A1367, where the center is assumed to be at the
mid-point between the two brightest ellipticals.  Systemic
heliocentric velocity and velocity dispersion are taken from the
literature, according to the references quoted in the
table. Heliocentric velocities are converted to the cosmic microwave
background (CMB) reference frame using the CMB dipole determintion
from Kogut \etal (1993\markcite{Kog93}).  Population fractions data
are from Dressler (1980\markcite{Dre80}), and Giovanelli \& Haynes
(1985\markcite{9clust}), while the fraction of HI deficient objects in
a cluster (this is the fraction of spiral galaxies that have HI
deficiency $>0.3$, according to the definition of Giovanelli \& Haynes
1985\markcite{9clust} and Haynes \& Giovanelli 1984\markcite{HGisol})
is taken from Giovanelli \& Haynes (1985\markcite{9clust}), and Magri
\etal (1988\markcite{Magri88}). X-ray data are taken from 
David \etal (1993\markcite{Dav93}).

The most extreme properties within our sample are those of the Coma
cluster, a rich, dynamically relaxed, and X-ray luminous cluster, that
exhibits strong morphological segregation and HI deficiency among the
spiral galaxy population.  Then we have medium richness clusters like
A2634 and A1367: the first is a dynamically evolved cluster (see
however Pinkney \etal 1993\markcite{Pin93} and Schindler \& Prieto
1997\markcite{Schind97} for a discussion on a possible recent merger
event in A2634), exhibiting clear morphological segregation and
pronounced HI deficiency among the spiral galaxy population, while the
second is an irregular cluster, with only moderate morphological
segregation and HI deficiency.  A262 is a transitional cluster, where
low X-ray luminosity and velocity dispersion, cool ICM temperature,
and high spiral fraction, all indicative of a dynamically unevolved
state, coexist with the presence of a massive dominant galaxy, which
is typical of more evolved systems (see for example Forman \& Jones
1982\markcite{Form82}, and Sarazin 1988\markcite{Sarazin}). The
cluster exhibits moderate morphological segregation and HI deficiency.
The low end of the cluster richness distribution is represented by the
Pegasus and Cancer clusters. Both are spiral rich, low X-ray
luminosity objects, that exhibit little morphological segregation and
marginal HI deficiency.  Here we refer to Cancer as the largest of the
separate groups that were identified by Bothun \etal
(1983\markcite{Both83}) over the area and redshift range traditionally
associated with the Cancer cluster. This is group A in the
nomenclature of Bothun \etal The rich groups associated with the
galaxies NGC 507 and NGC 383 are often considered as a single entity,
the so-called Pisces cluster. However Sakai, Giovanelli \& Wegner
(1994\markcite{Shoko94}) have clearly shown that the two can be
separated. In both groups the galaxy population is dominated by
elliptical and S0 galaxies, and HI deficiency is almost absent among
the spiral galaxy population.

\subsection{Cluster membership}

A discussion of the membership criteria adopted for the TF work on the
clusters in the present sample was presented by G97a (their section
8).  Here we adopt the same basic criteria, with the advantage that
the bulk of the early-type galaxy population is concentrated towards
the cluster centers, where the application of those criteria is easier
than at the cluster periphery.  Therefore we do not separate our FP
sample into certain and possible cluster members, as done in G97a,b, but
we define only {\it bona-fide} cluster members.  The photometric
sample presented here, however, includes also a number of galaxies for
which spectroscopic data is not available, considered likely cluster
members on the basis of their size and I band flux.

Whenever caustic lines in redshift space (Kaiser
1987\markcite{Kaiser87}, Reg{\"o}s \& Geller 1989\markcite{Reg89}, van
Haarlem 1992\markcite{vHa92}) are available for one of the clusters in
our sample, membership determination is based on those lines, as
estimated for $\Omega_\circ = 0.3$.  This is the case for Coma, where
caustic lines have been published by Reg{\"o}s \& Geller
(1989\markcite{Reg89}), van Haarlem (1992\markcite{vHa92}), and
Gavazzi, Randone \& Branchini (1995\markcite{Gav95}); A2634, where
they have been published by Scodeggio et al. (1995\markcite{S95}); and
A1367, where they have been published by Reg{\"o}s \& Geller
(1989\markcite{Reg89}), and Gavazzi \etal (1995\markcite{Gav95}). For
Coma and A1367 we use the caustic lines published by Reg{\"o}s \&
Geller (1989\markcite{Reg89}).

For A262, we use the detailed analysis of the spatial and velocity
distribution of the galaxies in the A262 area presented by Sakai \etal
(1994\markcite{Shoko94}).  Although they did not derive redshift space
caustic lines for this cluster, their work can be used to derive
reliable membership criteria, out to a radial distance of $2^\circ$
from the cluster center, beyond which the contamination from the
sparse population of the Perseus-Pisces Supercluster becomes
significant.  We define as A262 members all galaxies within $2^\circ$
from the cluster center, with $cz_{hel}$ within the interval
3500--6300 \kms.  Pegasus is well separated in redshift space from
foreground and background objects, and membership assignments are not
problematic, especially for the early-type population.  We define as
Pegasus members all galaxies within $2^\circ$ from the cluster center,
with $cz_{hel}$ within the interval 2700--5100 \kms.  For Cancer, we
use the membership table published by Bothun \etal
(1983\markcite{Both83}) for group A.  For the NGC 383 and NGC 507
groups, we use the analysis published by Sakai \etal
(1994\markcite{Shoko94}). Although redshift space caustic lines could
not be reliably derived, because of the small samples involved,
membership can be assigned confidently, thanks to the relative
under-density of the region in front and behind the two groups.  Here
we define as NGC 507 members all galaxies within $1.5^\circ$ from the
group center, with $cz_{hel}$ within the interval 3500--6700 \kms, and
as NGC 383 members all galaxies within $1.0^\circ$ from the group
center, with $cz_{hel}$ within the interval 3800--6500 \kms.

\section{Observations}

All photometric measurement have been derived from I-band CCD
images. The adoption of I band for photometry was mainly dictated by
the practical notion that the cluster images contain several galaxies
per frame, both of early- and late-type, and thus can be useful for
TF application, where the I band is favored for compatibility with
G97a,b.

Most of the observations were obtained during three runs with the 0.9m
telescope of the Kitt Peak National Observatory (KPNO). A small number
of observations in the cluster A2634 was obtained during observing
runs with the 1.3m telescope of the Michigan-Dartmouth-MIT (MDM)
Observatory (runs 1 and 3 in Table 2). Other observations in A2634
were obtained by KPNO staff with the 0.9m telescope during a period of
service observations (run 2 in Table 2).  Table 2 lists all observing
runs during which data were obtained for this project, and gives
parameters of the instrumental setups. The most significant
difference between the MDM and the KPNO observations is the much
smaller field of view of the 1.3m MDM telescope, when compared with
the 0.9m KPNO telescope.

All CCD frames were obtained with 600 seconds integration time. For
the brightest objects, a second frame with shorter integration
(200--300 seconds) was necessary to avoid saturation of the image in
the nuclear region of the galaxy. The median effective seeing
(measured as the FWHM of the stellar light profile) for the entire
data set is 1.6''. Figure 1 shows the distribution of effective seeing
for the galaxy observations. The actual atmospheric seeing might have
been smaller at times, but the pixel size with which the majority of
the observations were obtained limits the effective seeing to
$\geq$1.2''.  Given the large field of view of the KPNO 0.9m
telescope, and the large density of galaxies in the cluster
environments studied here, a complete mapping of the clusters' central
regions was obtained, with a mosaic of frames on a regularly spaced
grid. Only for the galaxies at the periphery of the various clusters
off-grid pointings were used.

\section{Data reduction}

\subsection{Basic reduction}

The data reduction process was performed entirely using a combination
of standard and specially written IRAF\footnote{IRAF (Image Reduction
and Analysis Facility) is distributed by the National Optical
Astronomy Observatories.}  procedures. All frames were
bias-subtracted, and flat-fielded using a ``superflat'' obtained from
a median-filtered combination of a large (n$>$40) number of frames.
At this point, for ease of manipulation, one small frame (512 x 512
pixels) for each target galaxy was extracted from the 2048 x 2048 KPNO
pixel frames. The 1024 x 1024 pixel MDM frames were kept as single
frames.  The sky background was subtracted from all frames, using the
average number of counts measured in 10--12 ``empty-sky'''
regions. The typical number of pixels over which this average was
computed is 30,000. The accuracy of the background determination can
be estimated from the comparison of the independent values subtracted
from different 512 x 512 pixel frames extracted from the same $2048
\times 2048$ pixel frame. This uncertainty is typically of 0.2\%.  All pixels
contaminated by the light of foreground stars or other galaxies, or by
cosmic rays hits were blanked and excluded from the final steps of
surface photometry.

For galaxies severely contaminated by the light of a nearby object
(either another galaxy or a foreground star), an iterative subtraction
procedure was used to obtain the individual light profiles. First, a
model was fitted to the brighter object, after blanking the fainter
one(s).  After subtracting such model from the original image, a model
was fitted to the fainter object, which again was subtracted from the
original image.  The procedure was iteratively applied until
convergence was reached, usually at the second iteration, producing
companion-subtracted light profiles, although the uncertainty on the
photometry of these objects is somewhat higher than for isolated
galaxies. This approach is especially necessary in the inner parts of
clusters, where the galaxy density is high.

\subsection{Photometric calibration}

One or two Landolt (1992\markcite{Lan}) fields, containing each one
from 5 to 7 stars, were repeatedly observed during each night, at
airmasses between 1.2 and 2.5, to provide estimates of the photometric
zero-point and atmospheric extinction coefficient. Observations at I
and R band were used to obtain color corrected photometric solutions.
Typical uncertainties in the zero-point determination are 0.02 mag,
both at I and R band. All galaxy magnitudes were computed assuming a
constant R--I color of 0.70 for elliptical, S0, and S0a galaxies (see,
for example, Fukugita, Shimasaku \& Ichikawa
1995\markcite{Fuk95}). The uncertainty introduced by this assumption
is negligible, because the color term coefficient is always very small
($<$ 0.05). Galaxy observations were all obtained at airmass $<1.3$,
thereby minimizing the effects of atmospheric extinction.

Most observations were obtained in photometric conditions. Those that
were not were calibrated using either short exposure frames obtained
in photometric conditions (run 2 was calibrated entirely in this way,
using frames obtained during run 3), or taking advantage of the
partial overlap among frames in the mapping grid (non photometric
nights in runs 4, 5, 7, 8 were calibrated using this technique). In
either case 10--15 stars that are present in both the ``photometric''
and the ``non photometric'' image were selected, and instrumental
(i.e. non calibrated) magnitudes for all them were obtained in both
images.  After rejecting the obvious variable stars, mean magnitude
differences between the stars in the two images were used to calibrate
the ``non photometric'' frames.  The typical uncertainty of this
process is $\simeq$0.03 mag.

\subsection{Surface photometry}

Surface photometry was obtained using the GALPHOT software package,
written for IRAF/STSDAS\footnote{STSDAS (Space Telescope Science Data
Analysis System) is distributed by the Space Telescope Science
Institute, which is operated by AURA, under contract to the National
Aeronautics and Space Administration.} mainly by W. Freudling,
J. Salzer and M. Haynes.  A galaxy's light distribution was fitted
with elliptical isophotes, using a modified version of the STSDAS {\it
Isophote} package. The center, ellipticity and position angle of the
ellipse were free parameters, while the ellipse semi-major axis was
incremented by a fixed fraction of its value at each step of the
fitting procedure.  The procedure was stopped when the fitted mean
number of counts per pixel dropped below the rms dispersion around the
mean value. For this dataset, this typically happens close to the 23.5
mag arcsec$^{-2}$ isophote. The fitted parameters yielded a model of
the galaxy, which was used to compute integrated magnitudes as a
function of semi-major axis. At this point the measured fluxes in
counts per pixel were transformed into magnitudes per square arcsec,
and all magnitudes were calibrated, color-corrected, and corrected
for atmospheric extinction.

The radial surface brightness profiles were fitted with a $r^{1/4}$
profile (de Vaucouleurs 1948\markcite{Devauc48}), to determine the
effective radius $r_e$ (the radius that contains half of the galaxy
total light) and the effective surface brightness $\mu_e$ (the mean
surface brightness within the effective radius).  For all galaxies,
the fit was performed from an inner radius twice as large as the
seeing disk radius of the relative image. The outer radius was chosen
depending on the shape of the surface brightness profile. For those
objects that are well described by the $r^{1/4}$ law at all radii, the
fit was extended to the outermost isophotes. For all other galaxies,
only the central core was fitted.  The effective radius $r_e$ and the
surface brightness at $r_e$ were the two free parameters in the fit,
and the effective surface brightness $\mu_e$ was obtained from
$\mu(r_e)$ as $\mu_e =
\mu(r_e) - 1.392$ (this is strictly valid only for a galaxy that
follows the $r^{1/4}$ profile at all radii).  The median value for the
uncertainty of the fits is 6.2\% in $r_e$, and 0.07 mag arcsec$^{-2}$
in $\mu_e$.

Integrated fluxes within the faintest reliably measured isophote were
obtained. Total magnitudes were derived from those, by adding a term
which was obtained by extrapolating the surface brightness profile to
infinity.  For those galaxies not well represented by the $r^{1/4}$
law at all radii the outer portion of the profile was fitted using
either a different $r^{1/4}$ profile from the one used to fit the core
of the galaxy, or an exponential disk profile, depending on the
profile shape.  The median uncertainty with which total magnitudes
were derived is 0.06 mag.

Total magnitudes and effective surface brightnesses were corrected for
Galactic extinction and cosmological effects. Extinction corrections
were computed for each galaxy following the prescriptions of Burstein
\& Heiles (1978\markcite{BH}), using $A_I = 0.45 A_V$. These corrections 
range from 0.0 to 0.13 mag. Statistically equivalent results are
obtained using the extinction maps recently published by Schlegel,
Finkbeiner \& Davis (1998\markcite{Schleg98}).  Cosmological
K-corrections were computed assuming $F_\nu \sim \nu^0$ (the energy
distribution of early-type galaxies in the red is nearly flat) so that
$K_I = 2.5 \log (1+z)$, and range from 0.01 to 0.04 mag. The (1+z)$^4$
cosmological corrections to surface brightness range from 0.04 to 0.15
mag arcsec$^{-2}$.  Effective radius and effective surface brightness
were corrected for the smearing effects of seeing, following the
prescriptions of Saglia \etal (1993a\markcite{Sag93a}, see in
particular their Figure 8). The seeing correction produced a median
reduction of 2\% of the measured values of $r_e$, and a median
correction of 0.016 mag arcsec$^{-2}$ to the measured values of
$\mu_e$.

\section{The photometric sample}

In each cluster, surface photometry was obtained for E, S0, and S0a
galaxies, potentially good candidates for FP work. A few galaxies that
were classified as S0a/Sa or Sa were included in the sample because a
velocity dispersion measurement was already available for them.  Known
foreground or background objects were excluded, while a number of
galaxies without redshift measurement, but with flux and size that
make them likely cluster members, were instead included in the
samples, except in the case of the Cancer cluster, where foreground
and background contamination is too pronounced. Galaxies in close
pairs or multiplets, and galaxies contaminated by bright foreground
stars projected over the galaxy or in its immediate vicinity were
rejected, unless a velocity dispersion measurement was already
available. Only in these cases was an attempt made to subtract the
contribution of the contaminating object to the observed light
distribution (see section 4.1). In total, photometric properties for
636 galaxies were measured. The final samples are close to
magnitude-limited in nature, as will be seen in the discussion on the
FP relation (Scodeggio \etal 1998b\markcite{pap5}; see also the
discussion in Paper II).

The photometric data are presented in Table 3. Only the first page of
this table is presented here for guidance regarding its form and
content; the entire table is available in digital form on the Journal
Electronic Edition, or from the authors.  The table is organized as
follows: \newline 
Col. 1: Galaxy name: if the galaxy is listed in the
UGC catalog (Nilson 1973\markcite{UGC}), the UGC number is given; if
not, our internal coding number, from the private galaxy catalog of
Giovanelli \& Haynes referred to as the AGC, is given. \newline
Cols. 2 and 3: the NGC or IC number, if available, and the CGCG field
and ordinal number within that field, if the galaxy is listed in the
CGCG (Zwicky \etal\ 1963-1968\markcite{CGCG}). For galaxies in the 
Coma cluster that are not listed in either one of those catalogs, 
the Dressler number (Dressler 1980\markcite{Dre80}) is given, if 
available. \newline 
Cols. 4 and 5: Right Ascension and Declination, in 1950.0 epoch. \newline 
Col. 6: morphological type code, in the RC3 (de Vaucouleurs \etal\ 
1991\markcite{RC3}) scheme (-5 for E, -2 for S0, 0 for S0a). 
Morphological types were derived from joint inspection of the blue plates 
of the Palomar Observatory Sky Survey (POSS), and of the CCD images. \newline 
Col. 7: Heliocentric radial velocity V$_{hel}$, in \kms. \newline 
Col. 8: Measured effective radius $r_e$, in arcsec. \newline 
Col. 9: Effective radius corrected for seeing effects $r_e^c$, in arcsec. 
\newline
Col. 10: Uncertainty on the effective radius $\epsilon_r$, in
arcsec. \newline 
Col. 11: Corrected metric effective radius $R_e$, in kiloparsec, obtained 
assuming the galaxy is at the distance indicated by the cluster redshift, 
in the CMB reference frame. \newline 
Col. 12: Measured effective surface brightness $\mu_e$, in I band mag
arcsec$^{-2}$. \newline 
Col. 13: Corrected effective surface brightness $\mu_e^c$, in I band mag 
arcsec$^{-2}$. $\mu_e^c$ is corrected for the smearing effects of seeing, 
for extinction within our Galaxy, for the cosmological k-correction term, 
and for $(1+z)^4$ cosmological dimming. \newline 
Col. 14: Uncertainty on the effective surface brightness $\epsilon_{\mu}$, 
in mag arcsec$^{-2}$. \newline
Col. 15: Measured apparent I band magnitude m$_I$, obtained
extrapolating the surface brightness profile to infinity. \newline
Col. 16: Corrected apparent I band magnitude m$_I^c$, obtained from
the value listed in col. 15, and corrected for extinction within our
Galaxy, and for the cosmological k-correction term.  No internal
extinction correction was applied. \newline 
Col. 17: Uncertainty on the measured magnitude, $\epsilon_{m}$. \newline 
Col. 18: Absolute magnitude M$_I$, obtained assuming that the galaxy is at 
the distance indicated by the cluster redshift, in the CMB reference
frame. \newline 
Col. 19: Quality codes, as follows: \newline 
First digit = Photometric code: 1 = observation obtained in photometric
conditions, uncertainty in the zero-point determination $<$0.03 mag; 2
= observation obtained in photometric conditions, uncertainty in the
zero-point determination 0.03--0.05 mag; 6 = observation obtained in
non photometric conditions, uncertainty in the zero-point
determination $<$0.03 mag; 7 = observation obtained in non photometric
conditions, uncertainty in the zero-point determination 0.03--0.06
mag. \newline Second digit = Image quality code: 0 = image is OK; 1 =
small amount of contamination from nearby stars or other galaxies; 2 =
significant amount of contamination from nearby stars or other
galaxies; 3 = iterative subtraction of the contaminating object
required; 4 = flat-fielding of sub-standard quality, due to scattered
light from a nearby bright star. \newline Third digit = Profile code:
1 = $r^{1/4}$ fit good at all radii; 2 = $r^{1/4}$ fit acceptable at
all radii; 3 = $r^{1/4}$ fit good only in the core of the galaxy; 4 =
the $r^{1/4}$ fit does not provide an acceptable description of the
observed photometric profile. \newline If an asterisk appears in this
column, special comments on the object are included at the foot of the
table.

A small subset of these data has already been published in Paper I, 
limited to galaxies in the Coma and A2634 clusters. Since then, a few
galaxies have been reobserved, and the data combined to obtain new
measurements of the relevant photometric parameters. The data presented
here therefore supersede that previous report.

\section{Internal and external comparison}

Repeated observations of the same galaxies provide an opportunity to
evaluate the internal consistency of the photometric measurements.  In
total, repeated observations of 71 galaxies are available.  The
results of the comparison of effective radius, effective surface
brightness, and total magnitude are presented in Figure 2, and they
are listed in Table 4.  Also listed in this table are the mean
statistical uncertainties associated with the derivation of those
parameters, multiplied by a $\sqrt{2}$ factor to allow a direct
comparison with the rms scatter in the observations (assuming equal
error contributions to the rms scatter). Good agreement is found
between the separate measurements, with only few isolated significant
discrepancies. These are generally due to the uncertainties introduced
by the contamination from other nearby objects in the field, as
discussed in section 4.1.  There is also good general agreement between the
rms scatter derived from the comparison and the estimated statistical
uncertainties. 

The errors on $r_e$ and $\mu_e$ are highly correlated. Figure 3a shows
$\Delta \log R_e$ versus $\Delta \mu_e$ for the 71 pairs of
observations. The linear correlation coefficient between 
$\Delta \log R_e$ and $\Delta \mu_e$ is $R = 0.962$. 
Because $r_e$ and $\mu_e$ enter the FP via the combination 
$\log R_e - 0.32~ \mu_e$, the correlated nature of the errors 
on these two quantities translates
into a predicted uncertainty on their combination of 0.013, to be
compared with the observed scatter of 0.020 (see Figure 3b).  Note
however that a reduction of $R$ by 10\% (the approximate level of
uncertainty on its determination) would translate in an increased
predicted uncertainty on $\log R_e - 0.32~ \mu_e$ to 0.025.

No large dataset of I band photometric observations of early-type
galaxies is available for an external comparison of our photometric
measurements.  An indirect test can be however obtained using
comparisons with similar measurements presented by Lucey \etal\
(1991\markcite{Lucey91}), Saglia, Bender \& Dressler
(1993b\markcite{Sag93b}), J{\o}rgensen, Franx \& Kj{\ae}rgaard
(1995\markcite{Jor95}) for galaxies in the Coma cluster.  Since none
of these studies used I band for the photometric observations, it is
necessary to assume standard colors for E and S0 galaxies, before
comparisons can be made. The V band magnitudes from Lucey \etal\
(1991\markcite{Lucey91}), the B band magnitudes from Saglia \etal\
(1993b\markcite{Sag93b}), and the Gunn r band magnitudes from
J{\o}rgensen \etal\ (1995\markcite{Jor95}) have been transformed to I
band magnitudes using V--I, B--I, and r--I colors of 1.31, 2.25, and
1.07, respectively (Fukugita \etal\ 1995\markcite{Fuk95}), without
taking into consideration the small differences between the brightest
and the faintest objects introduced by the color-magnitude relation of
early-type galaxies. Because of this, the expected rms scatter in the
true colors of E and S0 galaxies around these mean values is expected
to be $\simeq0.2$ mag.

There are 57 galaxies in common with the sample of Lucey \etal\
(1991\markcite{Lucey91}), 47 with the sample of Saglia \etal\
(1993b\markcite{Sag93b}), and 106 with the sample of J{\o}rgensen
\etal\ (1995\markcite{Jor95}). Table 5 and the panels of Figure 4, 5, 
and 6 show the results of the comparisons of effective radius,
effective surface brightness, and the combination $\log R_e -
0.32~\mu_e$.  In these figures, the difference (us - literature) is
plotted vs. the average value we can derive for the various parameters
using the two respective measurements. There are significant
differences in the determinations of both the effective radius and the
effective surface brightness.  For example, differences as high as 2
mag in the determination of $\mu_e$, after the color correction
described above were applied, are present.  However it is possible to
show that the main source of these differences lies in the different
procedures adopted to perform the $r^{1/4}$ fit, and in particular, in
the choice of the portion of the photometric profile used for the
fit. In this work, the fit was extended to an outer radius chosen on
the basis of the shape of the photometric profile: for profiles that
do not follow the $r^{1/4}$ relation at all radii, only the core of
the galaxy was fitted. If the fit were to have been performed using
all the points in the photometric profile, the fitted profile would
have a different slope, and both the effective radius and the
effective surface brightness would have different values.  Another
possible source of discrepancy is the choice of the inner radius from
where to start the $r^{1/4}$ fit. In this work, this radius was chosen
to be twice the radius of the seeing disk. A different choice would
likewise lead to a fitted profile with a somewhat different slope,
especially for objects with small effective radius.

To quantify this effect, we have reanalyzed the galaxies that show a
discrepancy $|\Delta \mu_e| > 1$ mag in Figure 4 and 6 (a total of 20
galaxies). Considering these objects only, we observe differences
between our measurements and those taken from the literature of
$\Delta \log R_e = 0.48 \pm 0.33$ and $\Delta \mu_e = 1.46 \pm 1.13$,
respectively.  We have then redetermined $r_e$ and $\mu_e$ for all
these galaxies without using the $r^{1/4}$ fit. To do so, we measured
on the observed photometric profile the radius that includes half of
the total galaxy flux (as derived from the total magnitude, which, in
turn, was obtained extrapolating to infinity the outer parts of the
photometric profile), and the corresponding average surface
brightness.  The values of $r_e$ and $\mu_e$ determined using this
procedure are in much better agreement with those published by Saglia
\etal\ (1993b\markcite{Sag93b}), and by J{\o}rgensen \etal\ 
(1995\markcite{Jor95}): the new differences between our modified
measurements and those taken from the literature are $\Delta \log R_e
= 0.063 \pm 0.084$ and $\Delta \mu_e = 0.086 \pm 0.29$, respectively.
To help visualize the extent of this problem with different fitting
techniques, the panels of Figure 4, 5, and 6 show with filled symbols
the galaxies whose photometric profiles follow the $r^{1/4}$ relation
at all radii, and with unfilled symbols those galaxies that do not.
It is clear that the largest discrepancies are mostly associated with
the latter objects.

As we have shown, the observed differences in $\log R_e$ and $\mu_e$
originate from differences in the $r^{1/4}$ fit. As a result there is
a correlation between $\Delta \log R_e$ and $\Delta \mu_e$ that is
completely analogous to the correlation shown by the errors in $\log
R_e$ and $\mu_e$. Therefore, despite the large differences in the two
separate values of $\log R_e$ and $\mu_e$, we observe an excellent
agreement in the comparison of the combinations $\log R_e -
0.32~\mu_e$.  It is therefore possible to say that these external
comparisons show good overall agreement, and establish the good
quality of the photometric measurements presented in this work, as
well as those presented in the comparison samples.

\acknowledgments

We would like to thank Gary Wegner, Shoko Sakai, and Bob Barr for
their help with the MDM observations, and Bill Schoening for his
technical support during the various observing runs at Kitt Peak.
This research has made use of the NASA/IPAC Extragalactic Database
(NED) which is operated by the Jet Propulsion Laboratory, California
Institute of Technology, under contract with the National Aereonautics
and Space Administration. This research is part of the Ph.D. Thesis
of MS, and is supported by the NSF grants AST94--20505 and
AST96--17069 to RG and AST95--28860 to MPH.

%\newpage

\begin{table}
\tablenum{2}
\caption[]{Imaging observing runs}
{\small
\begin{tabular}{lccc}
%\multicolumn{4}{c}{\large Table 2 Imaging observing runs} \\
\\
\tableline 
\tableline
\\
 & Run 1 & Run 2, 4--8 & Run 3 \\
\\
\tableline
\\       
Telescope      & MDM 1.3m            & KPNO 0.9m       & MDM 1.3m       \\     
CCD            & Loral2048 (Wilbur)  & T2KA            & Tek1024 (Charlotte)   \\ 
Pixels         & 1024 x 1024         & 2048 x 2048     & 1024 x 1024   \\
Read-out noise & 6.0 e$^-$           & 4.0  e$^-$      & 4.8 e$^-$       \\
Gain           & 1.9 e$^-$/DU        & 3.2 e$^-$/DU    & 3.45 e$^-$/DU  \\
Pixel scale    & 0.63  ''/pixel      & 0.68 ''/pixel   & 0.51  ''/pixel    \\
Field of view  & 11 x 11 arcmin      & 23 x 23 arcmin  & 9 x 9 arcmin    \\
\\
\end{tabular}
\begin{tabular}{lcccc}
\tableline
\tableline 
\\
&  Run 1                & Run 2              &  Run 3      &  Run 4        \\
\\
\tableline  
\\
Dates     & Sep. 23-25 1992\tablenotemark{a}  & Aug. 18-20 1993\tablenotemark{b} & 
Sep. 8-9 1993\tablenotemark{a} & Apr. 14-19 1994   \\   
Photometric nights  & 3       & 0     & 2       & 3    \\
Marginal nights     & 0       & 3     & 0       & 1    \\
Nr. of galaxies     & 10      & 78    & 6       & 322   \\
\\
\tableline
\tableline
\\
      &  Run 5     &  Run 6   & Run 7   & Run 8  \\
\\
\tableline  
\\
Dates    & Aug. 31 - Sep. 6 1994 & Oct. 8 1994\tablenotemark{a} & 
Mar. 6 1995\tablenotemark{a}  & Sep. 24 - Oct. 1 1995    \\
Photometric nights & 3        & 1        & 0        & 2  \\
Marginal nights    & 1        & 0        & 1        & 2  \\
Nr. of galaxies    & 223      & 16       & 5        & 47 \\
\\
\tableline
\tableline
\end{tabular}

\tablenotetext{a}{Observing run mainly dedicated to a different
project; only a small portion of telescope time dedicated to this project.}
\tablenotetext{b}{Service observing by KPNO personnel.}
}
\end{table}

\begin{deluxetable}{lcrrr}
\tablewidth{0pt}
\tablenum{4}
\tablecaption{Imaging data quality: internal comparison}
\tablehead{
\colhead{Parameter} & \colhead{N$_{gal}$} & \colhead{mean} & \colhead{rms} & 
\colhead{mean}  \nl
&& \colhead{difference} & \colhead{scatter} & \colhead{uncertainty}
}
\startdata
$\log r_e$             &  71 & -0.003\phm{aa} &  0.044\phn  &  0.038\phm{aa}   \nl
$\mu_e$                &  71 & -0.001\phm{aa} &  0.131\phn  &  0.113\phm{aa}   \nl
$m_{tot}$              &  71 & 0.071\phm{aa}  &  0.087\phn  &  0.084\phm{aa}   \nl
$\log r_e -0.32 \mu_e$ &  71 & 0.001\phm{aa}  &  0.020\phn  &  0.013\phm{aa}   \nl
\enddata
\end{deluxetable}

\begin{deluxetable}{lcrr}
\tablewidth{0pt}
\tablenum{5}
\tablecaption{Imaging data quality: External Comparison}
\tablehead{
\colhead{Parameter} & \colhead{N$_{gal}$} & \colhead{mean} & \colhead{rms} \nl 
&& \colhead{difference} & \colhead{scatter}
}
\startdata   
& \multicolumn{3}{c}{Saglia et al. (B band)} \nl
\nl
$\log r_e$ &  47 &  0.158\phm{aa}  &      0.240\phn \nl
$\mu_e$    &  47 &  0.365\phm{aa}  &      0.813\phn \nl
$\log r_e-0.32 \mu_e$ & 47 &  0.034\phm{aa}   &  0.043\phn \nl
\nl
& \multicolumn{3}{c}{Lucey et al. (V band)} \nl
\nl
$\log r_e$  & 57 &  0.030\phm{aa}  &   0.100\phn \nl
$\mu_e$     & 57 &  -0.100\phm{aa} &   0.353\phn \nl
$\log r_e-0.32 \mu_e$ & 57 &  0.060\phm{aa}  &  0.030\phn \nl
\nl
& \multicolumn{3}{c}{J{\o}rgensen et al. (r band)} \nl
\nl
$\log r_e$ & 106 &   0.173\phm{aa}  &    0.248\phn \nl
$\mu_e $   &  106 &   0.438\phm{aa}  &    0.741\phn \nl
$\log r_e-0.32 \mu_e$ & 106  &  0.031\phm{aa}  &   0.038\phn \nl
\enddata
\end{deluxetable}

\newpage

\plotone{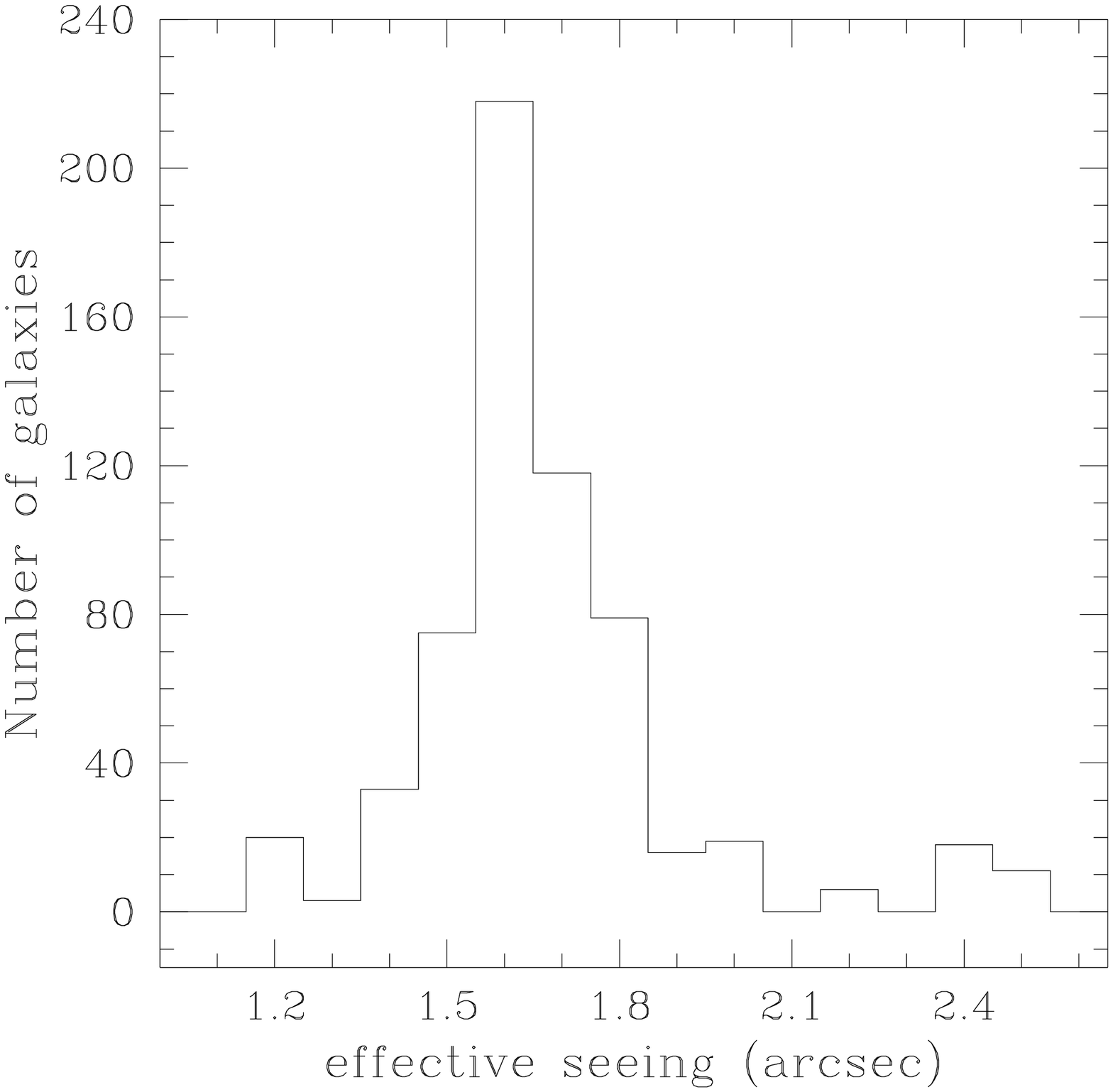}
\figcaption[fp_univers1_fig1.ps]{The distribution of effective seeing for 
the galaxy observations.  The pixel size of 0.68 arcsec. limits this
effective seeing to be no smaller than $\simeq$1.2 arcsec.}
\newpage

\plotone{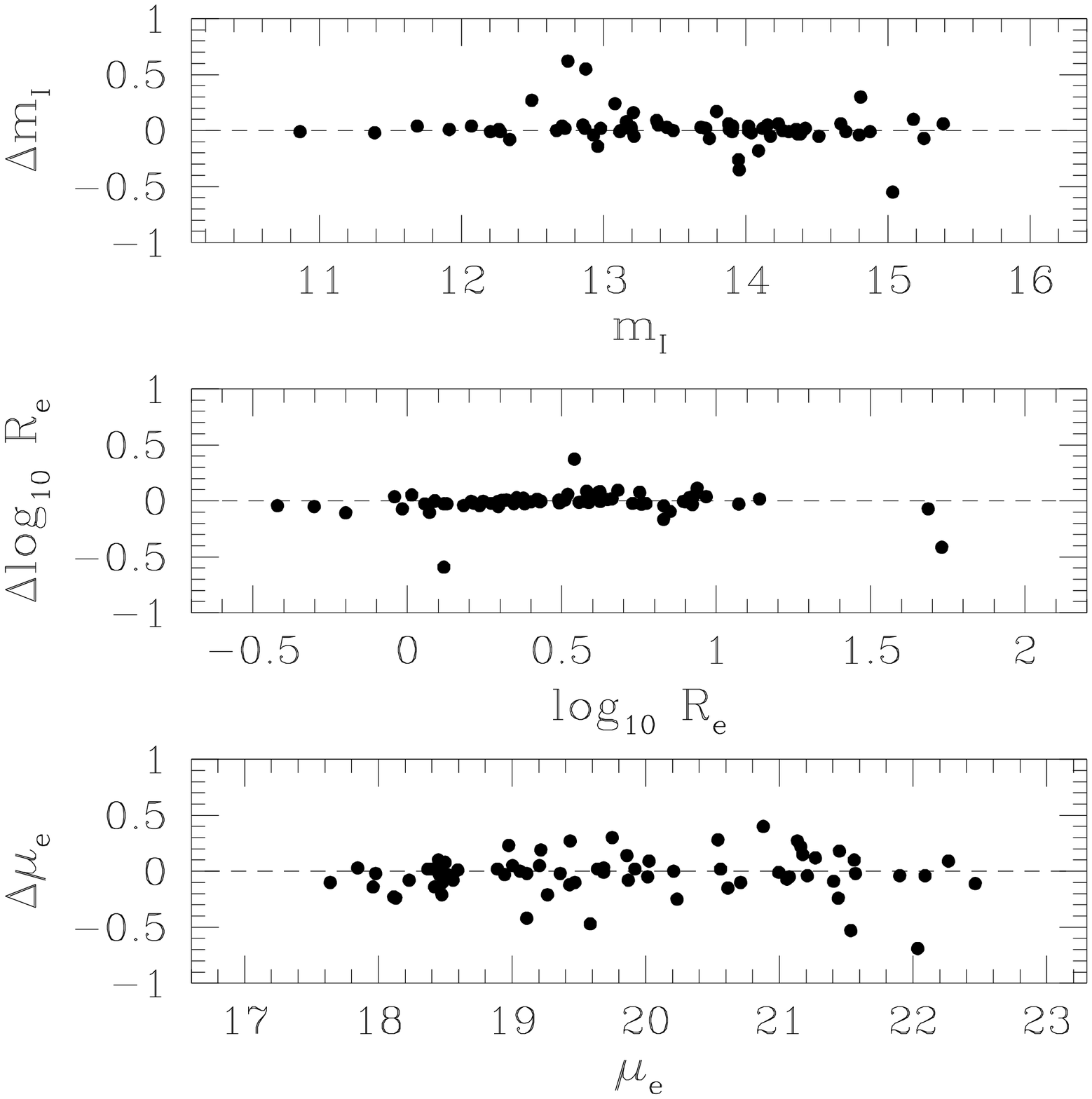}
\figcaption[fp_univers1_fig2.ps]{Internal comparison of the global photometric
parameters. $\Delta m_I$, $\Delta log_{10} R_e$, and $\Delta \mu_e$
are the differences between the values of the total magnitude,
effective radius, and effective surface brightness obtained from
repeated observations of 71 galaxies. They are plotted as a function
of the mean value of the relative parameter, obtained combining the
different measurements.}

\plotone{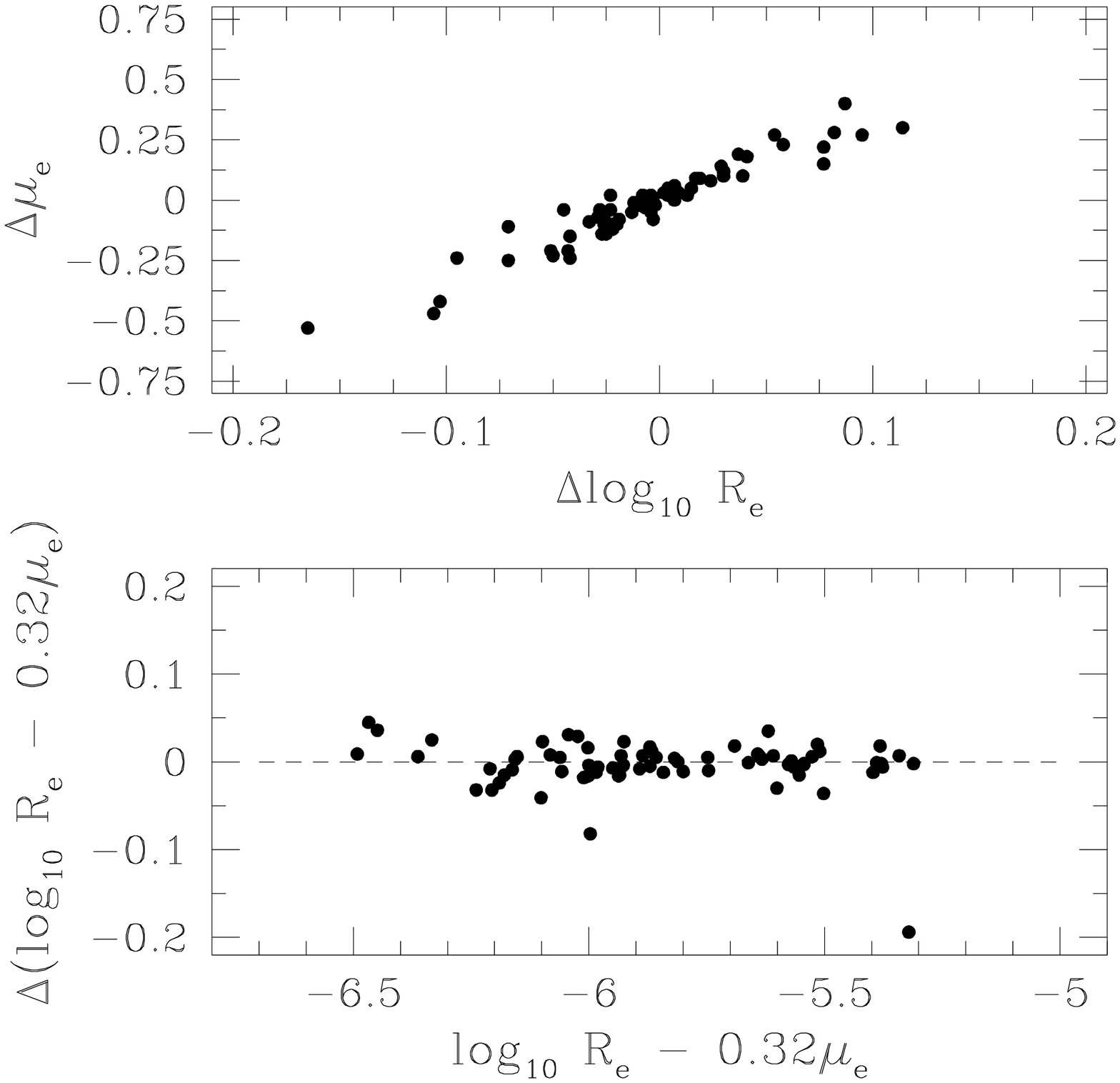}
\figcaption[fp_univers1_fig3.ps]{(a) The correlation of the errors in 
$\log r_e$ and $\mu_e$.  The figure shows the differences measured
from repeated observations.  The linear correlation coefficient for
this data-set is 0.962.  (b) Internal comparison of the combination
$\log r_e - 0.32~\mu_e$, which is the combination that enters in the
FP relation. As in Fig.2, the difference derived from repeated
observations is plotted against the mean value of the parameter.}

\plotone{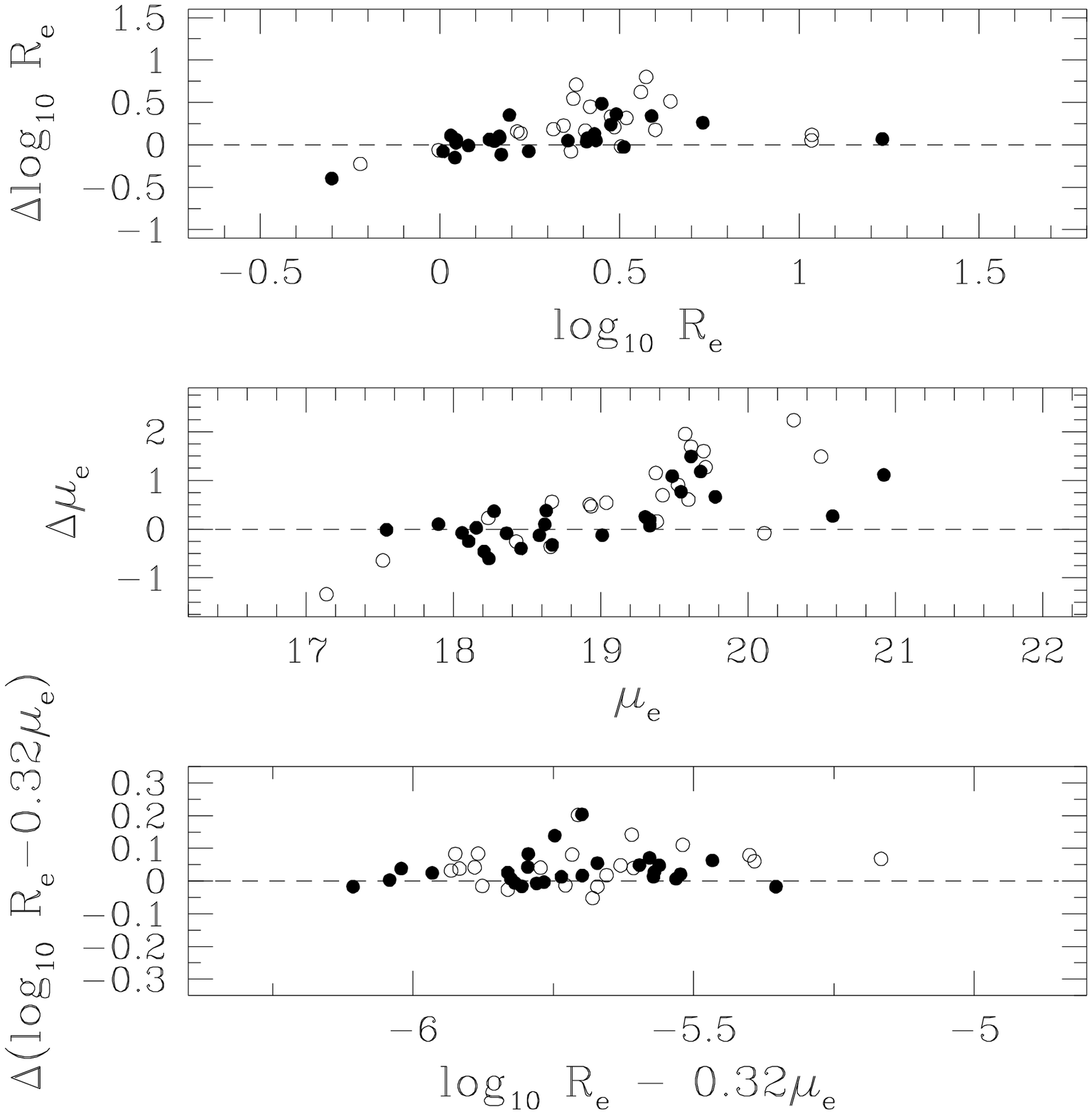}
\figcaption[fp_univers1_fig4.ps]{External comparison of the global photometric 
parameters for galaxies in the Coma cluster with the B band data from
Saglia et al.  B magnitudes were converted to I magnitudes for the
comparison, assuming a fixed B-I color (see text). Filled symbols
identify the galaxies whose photometric profile follows the $r^{1/4}$
relation at all radii, while empty symbols identify those galaxies
that deviate significantly from such a profile. The differences (us -
Saglia) are plotted against the mean value of the different
parameters, obtained combining the two measurements.}

\plotone{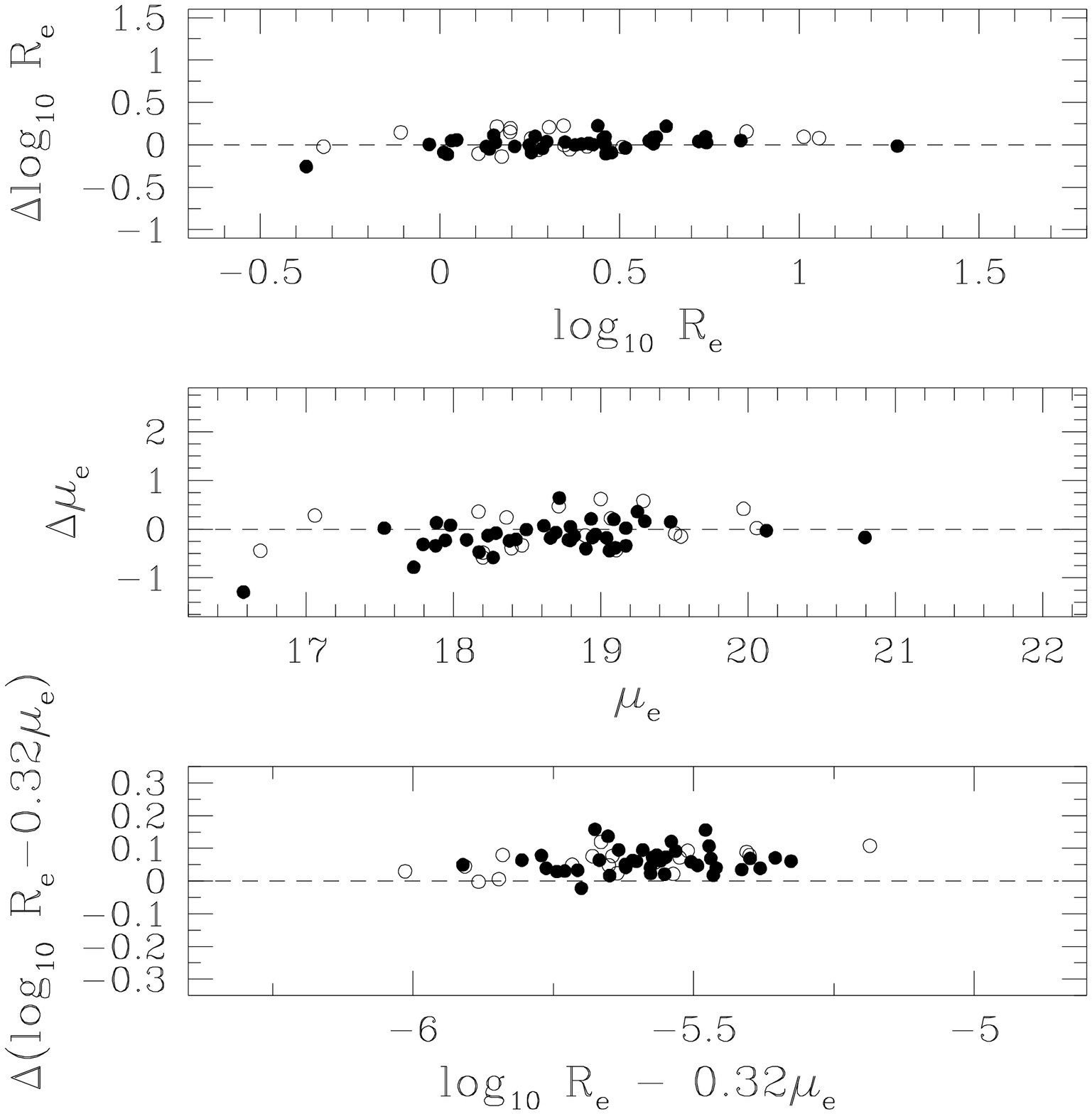}
\figcaption[fp_univers1_fig5.ps]{External comparison of the global photometric 
parameters for galaxies in the Coma cluster with the V band data from
Lucey et al.  V magnitudes were converted to I magnitudes for the
comparison, assuming a fixed V-I color (see text). Filled symbols
identify the galaxies whose photometric profile follows the $r^{1/4}$
relation at all radii, while empty symbols identify those galaxies
that deviate significantly from such a profile. As in the previos
figure, differences (us - Lucey) are plotted against the mean value of
the different parameters.}

\plotone{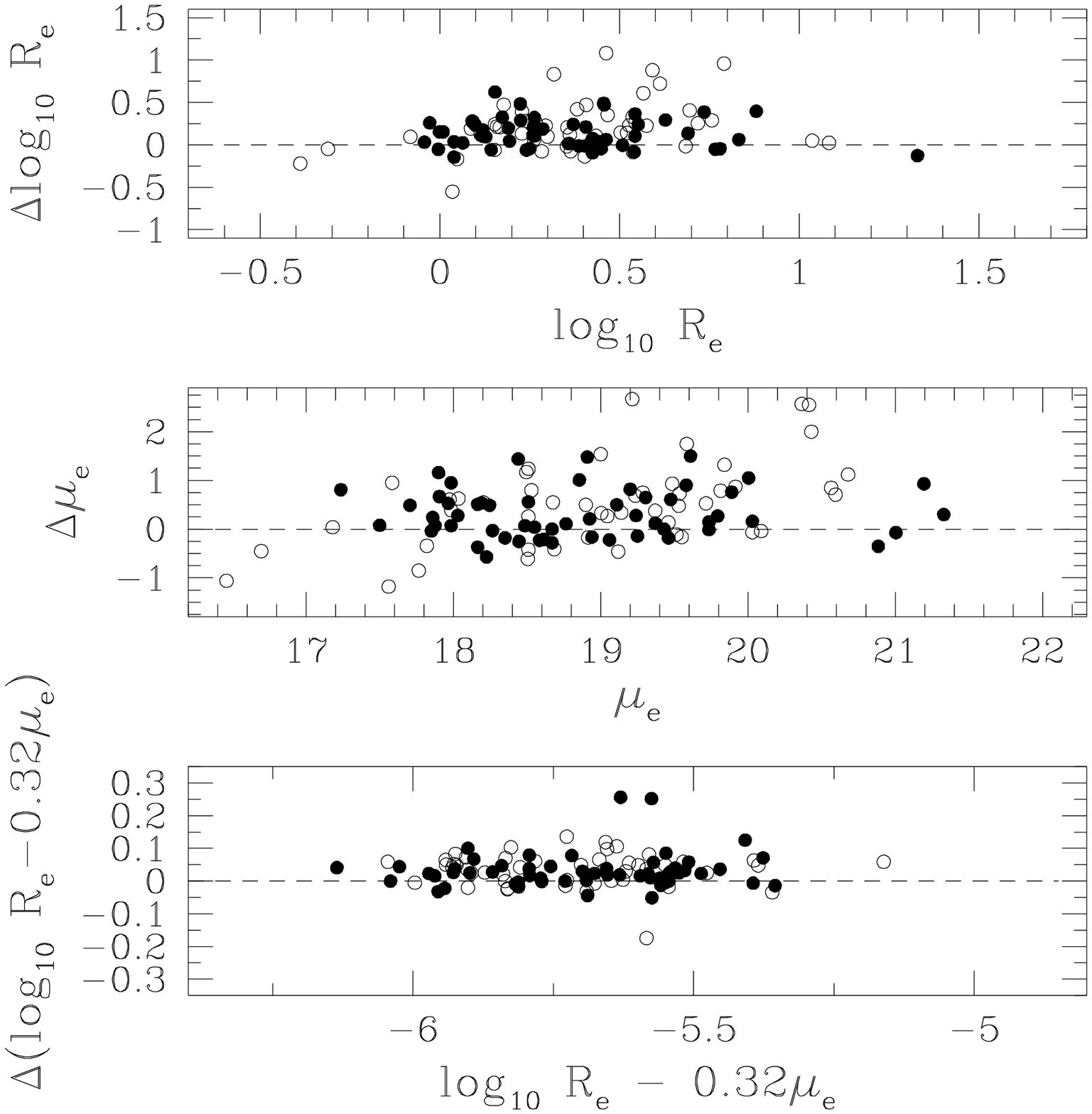}
\figcaption[fp_univers1_fig6.ps]{External comparison of the global photometric 
parameters for galaxies in the Coma cluster with the r band data of
J{\o}rgensen et al. r magnitudes were converted to I magnitudes for
the comparison, assuming a fixed r-I color (see text). Filled symbols
identify the galaxies whose photometric profile follows the $r^{1/4}$
relation at all radii, while empty symbols identify those galaxies
that deviate significantly from such a profile. As in the previos
figure, differences (us - J{\o}rgensen) are plotted against the mean
value of the different parameters.}

\end{document}

%\begin{center}
\ptlandscape

%\end{center}